\documentclass[a4paper]{article}
\usepackage{ASVspoof}
\usepackage{epsfig,amssymb,amsmath}
\usepackage{authblk}
\usepackage{hyperref}
\usepackage{xcolor}
\ninept

\newcommand{\mat}[1]{\mathbf{#1}}

\title{AASIST3: KAN-Enhanced AASIST Speech Deepfake Detection using SSL Features and Additional Regularization for the ASVspoof 2024 Challenge}

\makeatletter
\def\name#1{\gdef\@name{#1\newline\vspace{-1em}}}
\makeatother

\name{{\em Kirill Borodin$^{2,*}$, Vasiliy Kudryavtsev$^{2,*}$, Dmitrii Korzh $^{1,3,*}$,  Alexey Efimenko$^{2,*, \dagger}$}\newline
     {\em Grach Mkrtchian $^{2}$, Mikhail Gorodnichev $^{2}$, Oleg Y. Rogov$^{1,3}$}
     }

\address{\qquad $^1$AIRI, $^2$MTUCI, $^3$Skoltech, Moscow, Russia \newline
{\small \tt korzh@airi.net}}

\begin{document}

\maketitle

\begingroup
\renewcommand\thefootnote{\relax}
\footnotetext{* Equal contribution.}
\addtocounter{footnote}{-1}
\footnotetext{† Contributed during internship in AIRI.}
\addtocounter{footnote}{-1}
\endgroup


%
\begin{abstract}
Automatic Speaker Verification (ASV) systems, which identify speakers based on their voice characteristics, have numerous applications, such as user authentication in financial transactions, exclusive access control in smart devices, and forensic fraud detection. However, the advancement of deep learning algorithms has enabled the generation of synthetic audio through Text-to-Speech (TTS) and Voice Conversion (VC) systems, exposing ASV systems to potential vulnerabilities. To counteract this, we propose a novel architecture named AASIST3. By enhancing the existing AASIST framework with Kolmogorov-Arnold networks, additional layers, encoders, and pre-emphasis techniques, AASIST3 achieves a more than twofold improvement in performance. It demonstrates minDCF results of 0.5357 in the closed condition and 0.1414 in the open condition, significantly enhancing the detection of synthetic voices and improving ASV security.

\textbf{The new version of the model is publicly available at \href{https://huggingface.co/lab260/Spectra-AASIST3}{\underline{HuggingFace (2026)}}}


\end{abstract}

\section{Introduction}
Automatic Speaker Verification (ASV) systems are designed to identify speakers based on their voice characteristics. These systems have a variety of applications, including in the financial sector for user authentication during transactions, in smart devices to ensure exclusive access for the owner to control their equipment, and in forensic analysis to detect fraud cases.

However, the advent of deep learning algorithms has rendered ASV systems susceptible to many assaults. The public accessibility of Text-to-Speech (TTS) and Voice Conversion (VC) systems with pre-trained weights permits any user with access to computational resources, including cloud GPUs, to refine these models for potentially malevolent objectives.

To effectively counter such attacks, the implementation of anti-spoofing systems is imperative. The ASVSpoof community is engaged in active research in this field, as evidenced by the compilation of diverse data corpora for the development of both countermeasure (CM) systems and spoofing-aware speaker verification (SASV) systems. This research is documented in the following publications: \cite{asvspoof13, asvspoof15, asvspoof17, asvspoof19, asvspoof21}. Moreover, the Singfake dataset \cite{singfake} was developed to detect AI-generated vocals within the musical domain. Additionally, the SVDD project has significantly contributed to advancing voice spoofing countermeasures.


Today, many techniques are utilized to detect voice spoofing, including those based on Convolutional Neural Networks (CNN) \cite{lavrentyeva19_interspeech,choi22c_interspeech}, ResNet-like architectures \cite{alzantot19_interspeech,lai19b_interspeech,castan22_odyssey, Kwak_artcle}, Time Delay Neural Networks (TDNN) \cite{chen21_asvspoof, Wu22}, and transformers \cite{Khan23}. The AASIST architecture \cite{aasist} has demonstrated particular robustness, as confirmed by numerous studies. Various modifications have been proposed to enhance the generalization capability of AASIST, including the use of a Res2Net encoder \cite{aassist2}, wav2vec \cite{w2v2_aasist}, fusion of different audio representations \cite{s2pecnet}, application of specific training schemes such as SAM \cite{sam}, ASAM \cite{asam}, SWL \cite{swl}, as well as the use of alternative loss functions \cite{GCE,ding2022samospeakerattractormulticenter}.

In the present study, we propose an innovative architecture, AASIST3, developed on an AASIST base to detect speech deepfakes. The main modifications include:
\begin{itemize}
\item The modification of attention layers in GAT, GraphPool, and HS-GAL utilizing KAN\cite{kan} is based on the primary PreLU activation function and learnable B-splines, allowing for the extraction of more relevant features.
\item The scaling of the model in width using the proposed KAN-GAL, KAN-GraphPool, and KAN-HS-GAL techniques enabled the extraction of more complex parameters, resulting in enhanced model performance.
\item The primary methods of data pre-processing employed were diverse augmentations and pre-emphasis, intending to obtain more meaningful discriminative frequency information.
\end{itemize}

\section{Preliminaries}
\subsection{Kolmogorov-Arnold Network}
The Kolmogorov-Arnold theorem \cite{kolmogorov1, kolmogorov2, kolmogorov3} postulates that any continuous multivariate function $f: [0,1]^n \rightarrow \mathbb{R}$ can be represented as a finite composition of continuous unary functions and a binary addition operation. More specifically:
\begin{equation}
    f(x) = f(x_1, x_2, ..., x_n) = \sum^{2n}_{q=0} \Phi_q \sum^{2n}_{p=1} \phi_{q,p}(x_p),
\end{equation}
where $\phi_{q,p}: [0,1] \rightarrow \mathbb{R}$ and $\Phi_q: \mathbb{R} \rightarrow \mathbb{R}$

As all the functions to be learned by the model are one-dimensional, Liu et al.\cite{kan} proposed that each 1D function be parameterized as a B-spline curve and a basis function:
\begin{equation}
    \phi(x) = w_bb(x)+w_s \operatorname{spline}(x),
\end{equation}
where $b(x)$ represents the local basis function(eq. \ref{eq3}), while $w_b$ and $w_s$ denote trainable parameters that have been initialized following Kaiming initialization.
\begin{equation}\label{eq3}
    b(x) = \operatorname{PReLU}(x) = \max(0,x) + a\cdot \min(0,x)
\end{equation}
where $a$ is the trainable parameter, $\operatorname{spline}(x)$ - linear combination of B-splines:
\begin{equation}
    \operatorname{spline}(x) = \sum^4_{i=0} c_i B_i(x)
\end{equation}
where $c_i$ represents the trainable parameter and $B_i$ denotes the unique spline. For each spline of order $\beta_d = 4$, a total of $G$ points were utilized:
\begin{equation}
    G = 2 \beta_d + \beta_N + 1,
\end{equation}
where grid size $\beta_N$ is set to 16. The points are located on the interval $[\theta_1, \theta_2]$ eq.\ref{eq6}, \ref{eq7}
\begin{equation}\label{eq6}
    \theta_1 = - \beta_d  h + \alpha_1
\end{equation}
\begin{equation}\label{eq7}
    \theta_2 = (N + \beta_d + 1) h + \alpha_1
\end{equation}
\begin{equation}
    h = \frac{\alpha_2 - \alpha_1}{\beta_N},
\end{equation}
where $[\alpha_1, \alpha_2]$ is the grid range. We set $\alpha_2=1$ and $\alpha_1=-1$.

A KAN layer with input dimensionality $n_{in}$ and output dimensionality $n_{out}$ can be represented as a matrix of one-dimensional functions:
\begin{equation}
    \Phi\{\phi_{q,p}\}, p=\{1,2,...,n_{in}\}, q=\{1,2,...,n_{out}\}.
\end{equation}
In matrix form, the KAN layer can be expressed as follows:
\begin{equation}\label{eq:kanforwardmatrix}
    \mat{x}_{l+1} = 
    \underbrace{\begin{pmatrix}
        \phi_{1,1}(\cdot) & \phi_{1,2}(\cdot) & \cdots & \phi_{1,n_{l}}(\cdot) \\
        \phi_{2,1}(\cdot) & \phi_{2,2}(\cdot) & \cdots & \phi_{2,n_{l}}(\cdot) \\
        \vdots & \vdots & & \vdots \\
        \phi_{n_{l+1},1}(\cdot) & \phi_{n_{l+1},2}(\cdot) & \cdots & \phi_{n_{l+1},n_{l}}(\cdot) \\
    \end{pmatrix}}_{\mat{\Phi}_l}
    \mat{x}_{l},
\end{equation}
Where $\Phi$ is the matrix of the function of the KAN layer. Thus, the KAN layer can be denoted as:
\begin{equation}
    \operatorname{KAN}(X)=\Phi X.
\end{equation}

\subsection{Audio preprocessing}
In light of the hypothesis that high frequencies facilitate the model's ability to differentiate between bona fide and spoof utterances, we employed a pre-emphasis technique on the input signal:
\begin{equation}
    x_l = x_l - 0.97 \cdot x_{l-1},
\end{equation}
where $l$ equals 1, 2, 3, .., L, L represents the length of the audio signal, and 0.97 is the pre-emphasis factor. The pre-emphasis process suppresses low and enhances high frequencies, facilitating the model's ability to focus on more relevant features specific to spoofing or bona fide utterances.

\subsection{SincConv frontend}
Following AASIST\cite{aasist}, we use the non-trainable SincConv\cite{sincconv} to extract features from preprocessed audio. SincConv applies the function $g(n, f_1, f_2)$ to the speech signal chunks $x(n)$ using the Hamming window function $w_n$:
\begin{equation}
    y(n) = x(n) \cdot g(n, f_1, f_2) * w(n)
\end{equation}
\begin{equation}
    g(n, f_1, f_2) = 2 f_2 \operatorname{sinc}(2 \pi f_2 n) - 2 f_1 \operatorname{sinc}(2 \pi f_1 n)
\end{equation}
\begin{equation}
    \operatorname{sinc}(x) = \frac{\sin(x)}{x}
\end{equation}
\begin{equation}
    w(n) = 0.54 - 0.46 \cos{\frac{2 \pi n}{L}},
\end{equation}
where $f_1,f_2$ are fixed parameters equal to the minimum and maximum possible frequencies in the Mel-spectrogram of the passed signal

\subsection{Wav2Vec2 frontend}
Wav2Vec2 \cite{w2v2}, developed by Facebook AI, is a state-of-the-art method for converting audio to text. This model utilizes the Transformer architecture first introduced in \cite{vaswani2017attention}.
The Wav2Vec2 \cite{w2v2} architecture consists of two main components:
Encoder Layer: This layer transforms the input audio data into a sequence of hidden states. It consists of convolutional layers that transform the input audio data into a sequence of hidden states.
Predictor Layer: This layer takes the sequence of hidden states from the encoder layer and predicts the next hidden state. It uses the Transformer architecture, which allows the model to consider context when predicting the next state.
A key feature of Wav2Vec2 is that it is trained unsupervised, meaning it does not require labeled data for training. Instead, it uses a method called contrastive learning to learn audio representations.
As a front-end component for a scientific paper, Wav2Vec2 can automatically transcribe audio to text, which can help analyze audio data or create text versions of audio recordings.

\subsection{Encoder}
The encoder comprises six convolutional units. Except for the initial unit, each subsequent unit comprises two convolution units. The initial unit, however, comprises a single convolution unit in conjunction with another unit. The convolution unit implements the following transformation:
\begin{equation}
    \operatorname{ConvUnit(x)} = \operatorname{Conv}(\operatorname{SELU}(\operatorname{BatchNorm}(x)))
\end{equation}

Each unit's input is added to the output following the second convolutional unit in that block and downsampled using a convolutional layer if necessary. Subsequently, MaxPooling is applied following this skip connection.

\subsection{KAN-GAL}
In our work, we were also inspired by AASIST, which is based on the premise that graphs are fully connected because it is impossible to determine the degree of importance of each node to a given task in advance. In contrast to RawGat \cite{rawgat}, the activation functions were not employed due to the novel utilization of KANs.

The initial operation is to apply a dropout with a probability of 0.2. Subsequently, the attention mask is obtained by node-wise multiplication(denoted as "$\times$") of the nodes $h,~h \in \mathbb{R}^{N, D}$ and $N$ -- the number of nodes, $D$ -- node dimensionality, and subsequent passing through the KAN layer. Following this, the hyperbolic tangent is applied. The resulting expression is then matrix multiplied by the attention weights $W_{att}$, which have been initialized using Xavier initialization. The resulting values are then divided by the temperature T, resulting in an attention map A consisting of the corresponding probabilities, which is obtained using the softmax function:
\begin{equation}
    A = \operatorname{softmax}
\left( \frac{\operatorname{tanh}(\operatorname{KAN}_1(h \times h))W_{att}}{T} \right).
\end{equation}

The resulting attention map is projected using KAN, and in parallel, it is multiplied by the matrix and projected. The resulting projections are then added together and normalized by batch:
\begin{equation}
    \operatorname{KAN-GAL}(h) = \operatorname{BatchNorm}(\operatorname{KAN}_2(Ah) + \operatorname{KAN}_3(h)).
\end{equation}

\subsection{KAN-GraphPool}
As described in the previous section, the initial operation is to apply a dropout, after which the resulting output passes through the KAN layer and is transformed by the sigmoid function, represented by the symbol $\sigma(\times)$. The resulting value is then multiplied elementwise (denoted as "$\odot$") by the original graph. The dimensionality is subsequently reduced using the rank function, which returns the k most significant nodes in the resulting graph:
\begin{equation}
    \text{KAN-GraphPool} = \operatorname{rank}((\sigma(KAN(h))\odot h),k).
\end{equation}

\subsection{KAN-HS-GAL}
The layer accepts three inputs: $h_t$, which has a node dimensionality of $D_t$ (temporal graph), $h_s$, which has a node dimensionality of $D_s$ (spatial graph), and $S$ (stack node). Input graphs are projected into another latent space using KAN layers to equalize their dimensions and then merged to form a fully connected heterogeneous graph $h_{st}$ with node dimensionality $D_{st}$. A dropout with a probability of 0.2 is then applied to the resulting graph:
\begin{equation}
    h_{st} = \operatorname{CONCAT}(\operatorname{KAN}_1(h_t), \operatorname{KAN}_2(h_s)).
\end{equation}

The primary attention map $A$ is derived by multiplying each node in the graph $h_{st}$ by every other node, with the projection through the KAN layer undergoing a hyperbolic tangent transformation:
\begin{equation}
    A = \operatorname{tanh}(\operatorname{KAN}_3(h_{st} \times h_{st})).
\end{equation}

To derive a secondary attention map $B$, the initial attention map is partitioned into four matrices in accordance with the threshold $D_t$, defined as the number of nodes in the underlying graph $h_s$. Subsequently, these segments are multiplied by the weights $W_{11}, W_{12}$ and $W_{22}$:
\begin{equation}
B =
\begin{cases}
    \sum^{D_t+D_s}_{m=1} A_{ijm}\cdot W_{11m}, & \forall i \leq D_t \text{ and }j \leq D_t \\
    \sum^{D_t+D_s}_{m=1} A_{ijm}\cdot W_{22m}, & \forall i \geq D_t \text{ and }j \geq D_t \\
    \sum^{D_t+D_s}_{m=1} A_{ijm}\cdot W_{12m}, & \text{otherwise}.
\end{cases}
\end{equation}

The matrix is then divided by the temperature value T and passed through the Softmax function, thereby obtaining a probability map:
\begin{equation}
    \hat{B} = \operatorname{softmax}\left(\frac{B}{T}\right).
\end{equation}

To produce the attention map for the stack node update, the heterogeneous graph $h_{st}$ is taken and multiplied by the Stack Node $S$ node-wise. The resulting graph is then projected through the KAN layer, passed through the tangent, and matrix multiplied by the weights $W_m$. The value obtained is subsequently divided by the temperature and passed through softmax:
\begin{equation}
    A_m = \operatorname{softmax}\left( \frac{\operatorname{tanh}(\operatorname{KAN}_4(h_{st} \odot S))}{T} \right ).
\end{equation}

Combining two projections obtained using KAN layers to update a stack node is necessary. These are the projection of the matrix-multiplied attention map $A_m$ and graph $h_{st}$ and the projection of the stack node:
\begin{equation}
    \hat{S} = \operatorname{KAN}_5(A_m h_{st}) + KAN_6(S)
\end{equation}

The update of $h_{st}$ is achieved by combining two projections derived from KAN layers: the projection of the matrix multiplied secondary attention map $\hat{B}$ and the heterogeneous graph $h_{st}$ and the projection of the graph $h_{st}$ itself. The expression obtained is then subjected to batch normalization:
\begin{equation}
    \widehat{h_{st}} = \operatorname{BatchNorm}(\operatorname{KAN}_7(\hat{B}h_{st})+ \operatorname{KAN}_8(h_{st})).
\end{equation}

The resulting heterogeneous graph is then divided back into two components by multiplication with the mask matrices $M_1$ and $M_2$:
\begin{equation}
    \widehat{h_t} = \widehat{h_{st}} M_t
\end{equation}
\begin{equation}
    M_t = 
    \begin{pmatrix}
        I_t \\
        0_s
    \end{pmatrix},  I_t \in \mathbb{R}^{N \times D_t},  0_s \in \mathbb{R}^{N \times D_s}
\end{equation}
\begin{equation}
    \widehat{h_s} = \widehat{h_{st} M_s}
\end{equation}
\begin{equation}
    M_s = 
    \begin{pmatrix}
        0_t \\
        I_s
    \end{pmatrix},  0_t \in \mathbb{R}^{N \times D_t},  I_s \in \mathbb{R}^{N \times D_s}.
\end{equation}

\subsection{Models Architectures}
\subsection{AASIST3}\label{aasist3}
In the closed condition, the front-end is SincConv, whereas, in the open condition, it is Wav2Vec2 XLS-R\cite{w2v2} with additional linear or convolutional layers, which maintains the dimensionality.

The application of Max Pooling, Batch Normalization, and SELU preceded the encoder:
\begin{equation}
    \hat{x} = \operatorname{Encoder}(\operatorname{SELU}(\operatorname{BatchNorm}(\operatorname{MaxPool}(x))),
\end{equation}
where $x$ is an input pre-emphasized audio.

Subsequently, the acquired features were divided into temporal and spatial components, after which positional embedding ($\operatorname{PE}$) was incorporated. In this manner, graphs were formed, which were subsequently passed through a KAN-GAL and a KAN-GraphPool:
\begin{equation}
    h_t = \text{KAN-GraphPool}(\operatorname{KAN-GAL}(\max_t(\operatorname{abs}(\hat{x}) + \operatorname{PE}_t)))
\end{equation}
\begin{equation}
    h_s = \text{KAN-GraphPool}(\operatorname{KAN-GAL}(\max_s(\operatorname{abs}(\hat{x}) + \operatorname{PE}_s))).
\end{equation}

The resulting graphs and the previously initialized stack node were passed in parallel through four branches. The initial step is to apply KAN-HS-GAL in each branch:
\begin{equation}
    \begin{pmatrix}
        \widehat{h_t}_2 \\
        \widehat{h_s}_2 \\
        \widehat{S}_2
    \end{pmatrix} =
    \operatorname{KAN-HS-GAL}
    \begin{pmatrix}
        \widehat{h_t}_1 \\
        \widehat{h_s}_1 \\
        \widehat{S}_1
    \end{pmatrix}.
\end{equation}

The graphs are then passed through KAN-GraphPool, and another KAN-HS-GAL is applied similarly:
\begin{equation}
    \begin{pmatrix}
        \widehat{h_t}_3 \\
        \widehat{h_s}_3 \\
        \widehat{S}_3
    \end{pmatrix} =
    \operatorname{KAN-HS-GAL}
    \begin{pmatrix}
        \operatorname{KAN-GraphPool}(\widehat{h_t}_2) \\
        \operatorname{KAN-GraphPool}(\widehat{h_s}_2) \\
        \widehat{S}_2
    \end{pmatrix}.
\end{equation}

To produce the final predictions, all previously obtained graphs and Stack Nodes are stacked:
\begin{align}
    & H_t = \widehat{h_t}_1 + \widehat{h_t}_2 + \widehat{h_t}_3
    \\
    & H_s = \widehat{h_s}_1 + \widehat{h_s}_2 + \widehat{h_s}_3
    \\
    & S_f = \widehat{S}_1 + \widehat{S}_2 + \widehat{S}_3.
\end{align}
    
    
    

A dropout with a probability of $0.2$ is applied to all obtained graphs and the Stack Node after four branches. Subsequently, for temporal and spatial graphs, the node-wise maximum $H^{\max}$ and mean $H^{\operatorname{mean}}$ are identified, as well as the maximum Stack node $S_f^{max}$. The resulting values pass through the dropout with a probability of 0.5 and are then concatenated into the final hidden layer L:

\begin{equation}
    L = \operatorname{CONCAT}(H^{\max}_t, H^{\operatorname{mean}}_t, H^{\max}_s, H^{\operatorname{mean}}_t, S_f^{\max})
\end{equation}
After $L$, a KAN layer returns logits for each class.

\subsection{Wav2Vec2-Conv-AASIST-KAN}
In addition to the proposed AASIST3, we utilized the pre-trained Wav2Vec2 encoder for the feature and proceeded with 1D convolutions to provide the AASIST model with a KAN classification layer. It was motivated by inductive biases from a pre-trained speech encoder on a large-scale dataset in self-supervised (SSL) training, which is preferable for the open-set condition.

\section{Experiments and Results}

\subsection{Description of final approaches}\label{final_res}

For the Closed Condition, we considered our described AASIST3 model. The model was constrained to accept only four seconds of audio as input, which proved insufficient for the models to achieve a deeper comprehension of the audio as a whole. To address this limitation, the audio was fed into the model in four-second parts sequentially with a two-second overlap between them. We applied pre-emphasis for all audios with no augmentations.

The optimal models were identified upon testing the models on a closed test subset: one with two branches and one with four branches. Additionally, based on the hypothesis that SWL can enhance the results, the model incorporating SWL was utilized. The predictions of these models were averaged.

\begin{table}[th]
\caption{\label{resulttable} {\it Final evaluation results of submitted prediction in closed and open condition CM track.}}
\vspace{2mm}
\centerline{
\begin{tabular}{|c|c|c|c|}
\hline
condition & model & t-DCF & EER \\
\hline  \hline
closed & AASIST3  & 0.5357 & 22.67 \\  
open & $\tilde f$ & 0.1414 & 4.89 \\
\hline
\end{tabular}}
\end{table}

As illustrated in the table \ref{tableofresults}, many of our modifications produced notably superior outcomes compared to AASIST, even during the validation phase. However, using various techniques to enhance the quality of anti-spoofing models proved ineffective, with all approaches ultimately resulting in a decline in the observed results.

For the Open Condition, to provide the final prediction or probability that given audio $x$ is bonafide, we averaged the predictions of two of our models trained differently to increase generalization ability.

\begin{equation}
    \tilde f = \frac{f_{1}^{'}(x) + f_{2}^{'} (x)}{2},
\end{equation}
where
\begin{align}
    & f_{1}^{'}(x) = \sum_i^m f_1(x_i)
    \\
    & f_{2}^{'}(x) = \sum_j^l f_2(x_j),
\end{align}
and $\{x_i\}_{i=1}^{m}$ and $\{x_j\}_{j=1}^{l}$ are some parts of the original audio $x$, specifically, sequential parts with intersection (for example, 0-4s and 3-7s audio intervals) as in the submission for Closed Condition.
The $f_1$ is AASIST3 but with a pre-trained Wav2Vec2 feature encoder. The $f_2$ is our second model Wav2Vec2+Conv+AASIST+KAN.

$f_1$ was trained similarly as in closed condition only on the provided training set, whereas $f_2$ was trained on the union of the given training set plus additional bonafide audios from Mozilla CommonVoice, train part if VoxCeleb2. For $f_2$ training, a combination of weighted Cross-Entropy, Focal \cite{lin2017focal}, and LibAUCM \cite{yang2022algorithmic, yuan2023libauc} losses with Adam optimizer. Augmentation methods such as RIR, environmental and Gaussian noises, VAD, and pitch shifting were randomly applied. Pre-emphasizing was used before augmentations.

\begin{figure*}
    \centering
    \includegraphics[width=1\linewidth]{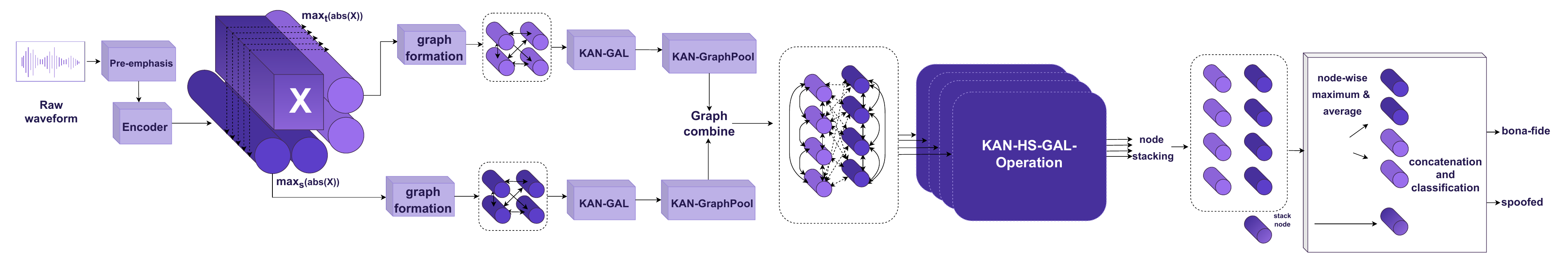}
    \caption{Architecture of the closed condition model.}
    \label{fig:enter-label}
\end{figure*}

\begin{figure*}
    \centering
    \includegraphics[width=0.7\linewidth]{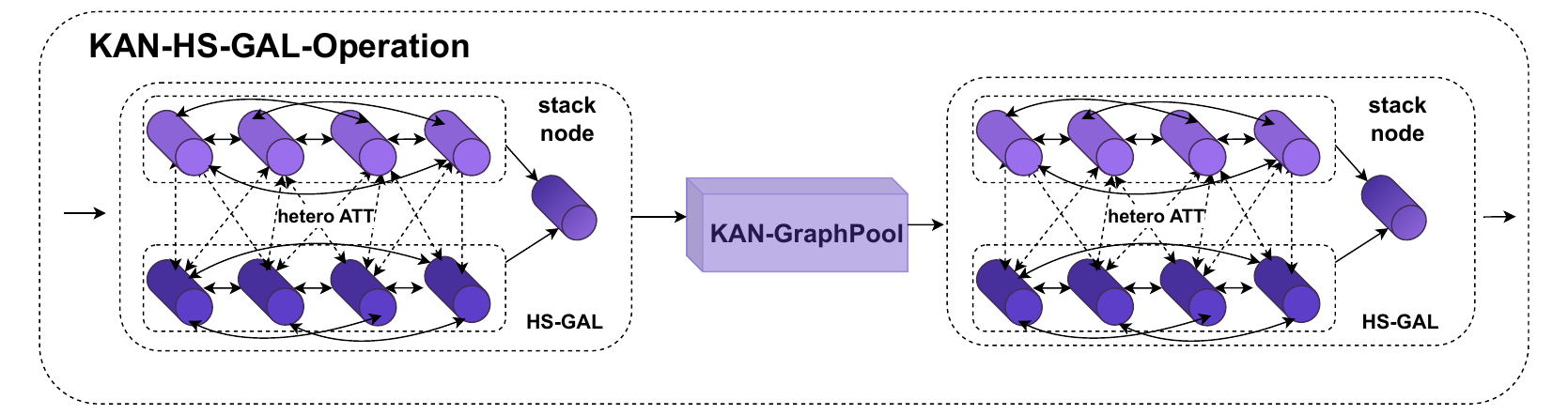}
    \caption{The KAN-HS-GAL Operation.}
    \label{fig:enter-label}
\end{figure*}

\subsection{Experiments with different frontends}\label{frontends}
Given the results presented by \cite{s2pecnet}, it was hypothesized that combining multiple representations might improve the result. Combining the raw waveform with the CQT and Mel-spectrograms was attempted, but no improvement was seen. In light of the findings in \cite{res2netwithf0}, we investigated using the f0 subband independently and in conjunction with SincConv. In addition, given the evidence presented in the study referenced in \cite{leaf}, which indicated that Leaf outperformed SincConv, we compared its performance with our model's. However, none of the modifications resulted in a performance improvement. For open conditions, the best result was shown by Wav2Vec2 \cite{w2v2} pre-trained on XLSR-300, as front-end based on a transformer and convolutional neural networks, which allows suitable encoding of both temporal and spatial information in audio. Also, experiments with XEUS \cite{chen2024towards} did not provide better results compared to other front-ends.


\subsection{Experiments with augmentations}\label{augmentations}
This study employed a series of augmentations, including pitch shift, speed change, Gaussian and environmental noise, different room impulse responses, harmonic and percussive components, stretching, voice activity detection (VAD) application, and pre-emphasis. A random portion of the audio sample was extracted or padded during training. In some experiments, the pre-emphasis was used as an augmentation, in others it was applied to each audio. Additionally, we employed Attention Augmented Convolutional Networks\cite{attenaugmentconv} and Rawboost\cite{rawboost}. The findings indicate that pre-emphasis on each audio track represents an effective approach. 


\subsection{Experiments with different base KAN functions}\label{ELU}
A comparison of AReLU\cite{arelu}, PReLU, SELU, and RReLU showed that PReLU gave the most robust results.


\subsection{Experiments with different KAN-based encoders}\label{KAN-encoders}

As one of the key ideas of our model is the use of KAN, we chose to explore the potential of KAN-based encoders, including KAN+tokenization\cite{ukan}, ReluConvKAN and WavKANConv\cite{drokin2024kolmogorovarnoldconvolutionsdesignprinciples}, both in stand-alone form and combined with a sharpness-aware minimisation\cite{sam} mechanism. When evaluated using a validation set with SAM, WavKanConv showed favorable results, table \ref{tableofresults}



Our experiments with different KAN encoders found that the model using the classic RawNet2-based encoder performed best.

\subsection{Experiments with different KANs}\label{polynomials}

A comparison was conducted between parametric B-spline functions and other types of polynomials, including Bessel polynomials, Chebyshev polynomials of the second kind, Gaussian radial basis functions, Fibonacci polynomials, radial-basis functions, and Jacobi polynomials. The results showed that the approximation with a B-spline of order 4 gave the most accurate results.


\subsection{Experiments with AASIST modification}\label{aasist_modification}

In order to test the effectiveness of the proposed methodology, several modifications were applied. These included the insertion of the third additional branch with channel-wise maximum, which was used to allow the extraction of more complex features. Furthermore, GraphPools were replaced by GALs to avoid removing a significant amount of information. The minimum was used instead of the maximum to form branches, and four branches with HS-GAL and SE\cite{se} in the encoder were used. Six branches were also used with HS-GAL, and positional encoding was introduced instead of positional embedding\cite{attentionisallyouneeded}. Finally, a comparison was made between the results obtained with two branches and those obtained with the proposed methodology. The results show that the best result is obtained using four branches with HS-GAL.


\subsection{Experiments with scaling of HS-GAL branches}\label{scaling_in_depth}

Additional HS-gals with different temperature values were applied to two branches. As shown in the table \ref{tableofresults},  the obtained results did not exhibit a considerably enhanced degree of improvement.



The results allow us to conclude that scaling the model in width is more optimal than scaling it in depth.

\subsection{Experiments with different encoders}\label{encoders}

Given that the original AASIST employs a RawNet2-based encoder, we postulated that a RawNet3-based encoder\cite{rawnet3} would enhance the model. Concurrently, we anticipate that S2pecNet\cite{s2pecnet}, as the authors have demonstrated, will improve the result through this integration of sound representations. Additionally, we explored the potential of WaveNet\cite{wavenet} as a front-end, but unfortunately, none of the experiments yielded a significant result. 


\subsubsection{Experiments with Res2Net-based encoders}\label{res2net}

Following AASIST2\cite{aassist2}, we attempted to utilize Res2Net in various configurations with disparate learning rates. Concurrently, we evaluated SR LA RES2net\cite{res2netwithf0} as a more sophisticated analog. The results of our investigation suggest that the application of Res2Net with the proposed AASIST configuration is not a viable approach.


\subsubsection{Experiments with ResNet-based encoders}\label{resnet}

Utilizing alternative encoders and modifications to the Res2Net encoder yielded no perceptible improvement in results. Therefore, an investigation was conducted into alternative changes to the ResNet encoder. These included the utilization of the f0 subband instead of the SincConv, the use of two ResNet encoders for different segments of audio, with and without RawBoost, the integration of ELA \cite{ela}, the substitution of BatchNorm with LayerNorm, the utilization of $\operatorname{prelu}$ as the activation function, and the integration of SE. The experimental results demonstrate that modifying the encoder does not improve outcomes.


\subsection{Experiments with different loss functions}\label{losses}

Furthermore, in line with AASIST2\cite{aassist2}, AM-Softmax and its predecessor, ArcFace\cite{arcface}, were also tested. Based upon the results of the study \cite{convnext}, focal loss was also tested. Furthermore, we attempted to utilize generalized cross entropy\cite{GCE}, the effectiveness of which has been previously established for AASIST. Additionally, as in the original AASIST, we attempted to utilize weighted cross-entropy. Finally, we selected multitask losses, hypothesizing that the model would be capable of extracting more complex features. However, our findings indicated that regular cross-entropy was indeed efficacious. Consequently, the deployment of loss to address class imbalance in our model can be considered ineffective. However, for better generalization, our second model was trained using a combination of weighted cross-entropy, focal loss, and LibAUCM \cite{yang2022algorithmic, yuan2023libauc} loss, which is implied for the x-risk minimization.


\subsection{Experiments with different optimizers}\label{optimizers}

The following optimizers were selected for our experiments: AdamW, Lion, NAdam, RAdam, and Adam. As illustrated in the table \ref{tableofresults}, the outcomes with AdamW and Lion exhibited a notable decline in performance. In addition, the results with RAdam were found to be unsatisfactory. The results on the development subset are presented in the table \ref{tableofresults}.



The findings indicate that Adam is the optimal optimizer for our model.

\subsection{Experiments with different learning methods}\label{learning_methods}

To enhance the generalization capacity, we employed a variety of techniques, including SAM \cite{sam}, ASAM \cite{asam}, and SWL \cite{swl}. These were initially utilized in the original paper about AASIST and led to a notable enhancement in the quality of the models. Furthermore, a cosine annealing scheduler and a weighted random sampler are employed. As illustrated in the table \ref{tableofresults}, these strategies have also been demonstrated to be ineffective.



It was found that none of the proposed learning methods improved the result.



\begin{table}[!htb]
\caption{\label{tableofresults} {\it Results of experiments with AASIST3 on dev subset.}}
\vspace{1mm}
\centerline{
\begin{tabular}{|c|c|c|}
\hline
Referenced & AASIST3 modification & t-DCF \\
\hline  \hline

\cite{aasist} &\textbf{Classic AASIST} & \textbf{0.5671} \\
\hline

sec. \ref{final_res} & \textbf{AASIST3} & \textbf{0.2657} \\
\hline

sec. \ref{frontends} &Leaf instead of SincConv & 0.52 \\
sec. \ref{frontends} &F0 subband + SincConv & 0.4406 \\
sec. \ref{frontends} &F0 subband instead of SincConv & 0.4225 \\
sec. \ref{frontends} &SincConv + CQT & 0.4083 \\
\hline
sec. \ref{augmentations} &AttentionAugmentedConv2d in encoder & 0.4762 \\
sec. \ref{augmentations} &Augmentations without pre-emphasis & 0.4495 \\
sec. \ref{augmentations} &All augmentations & 0.3624 \\
\hline

sec. \ref{ELU}&AReLU instead of PReLU in KAN & 0.3371 \\
sec. \ref{ELU} &SELU instead of PReLU in KAN  & 0.3295 \\
sec. \ref{ELU} &RReLU instead of PReLU in KAN & 0.3045 \\
\hline

sec. \ref{KAN-encoders} &ReLUConvKAN instead of Conv & 0.699 \\
sec. \ref{KAN-encoders} &UKAN in encoder & 0.3062 \\
sec. \ref{KAN-encoders} &WavKANConv in encoder & 0.3047 \\
sec. \ref{KAN-encoders} &WavKANConv in encoder + SAM & 0.2801 \\
\hline

sec. \ref{polynomials} &Bessel polynomials in KAN & 0.6805 \\
sec. \ref{polynomials} &Jacobi Polynomials in KAN & 0.5666 \\
sec. \ref{polynomials} &Legendre Polynomials in KAN & 0.4665 \\
sec. \ref{polynomials} &Gegenbauer polynomials in KAN & 0.4051 \\
sec. \ref{polynomials} &RBF in KAN & 0.3994 \\
sec. \ref{polynomials}&2nd Chebyshev polynomials & 0.3392 \\
sec. \ref{polynomials} &Fibonacci polynomials in KAN & 0.492 \\
\hline

sec. \ref{aasist_modification} &Utilising GATs instead of GraphPool & 0.4375 \\
sec. \ref{aasist_modification} &Gaussian RBF & 0.3536 \\
sec. \ref{aasist_modification} &4 branches of 2 HS-GALs + SE & 0.2905 \\
sec. \ref{aasist_modification} &3rd branch with channel-wise maximum & 0.4992 \\
\hline
sec. \ref{scaling_in_depth} &2 branches of 3 HS-GALs with temp=100 & 0.3077 \\
sec. \ref{scaling_in_depth} &2 branches of 3 HS-GALs with temp=150 & 0.2661 \\
sec. \ref{scaling_in_depth} &2 branches of 3 HS-GALs with temp=200 & 0.2862 \\
\hline

sec. \ref{encoders} &S$^2$pecNet with 40 batch size & 0.4291 \\
sec. \ref{encoders} &S$^2$pecNet with 28 batch size & 0.4225 \\
sec. \ref{encoders} &S$^2$pecNet with 20 batch size & 0.4185 \\
sec. \ref{encoders} &RawNet3 instead of RawNet2 & 0.4901 \\

sec. \ref{res2net} &Res2Net encoder with lr=1e-6 & 0.9066 \\
sec. \ref{res2net} &SR LA Res2Net encoder & 0.7203 \\
sec. \ref{res2net} &Res2Net encoder with lr=1e-5 & 0.6413 \\
sec. \ref{res2net} &SR LA Res2Net + f0 subband & 0.5463 \\
sec. \ref{res2net} &Res2Net encoder + PRELU with lr=1e-4 & 0.485 \\

sec. \ref{resnet} &LayerNorm instead of BatchNorm & 0.3591 \\
sec. \ref{resnet} &ResNet + effictive local attention & 0.3542 \\
sec. \ref{resnet} &ResNet with PReLU & 0.3216 \\
sec. \ref{resnet} &ResNet + SE & 0.2902 \\
\hline

sec. \ref{losses} &Generalized Cross Entropy Loss & 0.8438 \\
sec. \ref{losses} &ArcFace Loss & 0.4389 \\
sec. \ref{losses} &Multitask Loss & 0.3933 \\
sec. \ref{losses} & Focal Loss & 0.3489 \\
sec. \ref{losses} &AM Softmax & 0.3363 \\
\hline

sec. \ref{optimizers} &Lion instead of Adam & 0.3702 \\
sec. \ref{optimizers} &AdamW instead of Adam & 0.3200 \\
sec. \ref{optimizers} &NAdam instead of Adam & 0.2889 \\
sec. \ref{optimizers} &RAdam instead of Adam & 0.3006 \\
\hline
sec. \ref{learning_methods} &SAM rho=0.5 & 0.5012 \\
sec. \ref{learning_methods} &SAM rho=0.05 & 0.3727 \\
sec. \ref{learning_methods} &ASAM & 0.3532 \\
sec. \ref{learning_methods} &SWL & 0.2694 \\
sec. \ref{learning_methods} &Cosine annealing scheduler & 0.2991 \\
sec. \ref{learning_methods} &Weighted random sampler & 0.2989 \\


\hline
\end{tabular}}
\end{table}

\section{Conclusion}

The rapid development of various deep learning algorithms has created new opportunities for generating synthetic audio using TTS and VC systems. This progress, however, has introduced a corresponding vulnerability in ASV systems, necessitating the development of a CM system to detect synthetic voices. In this paper, we proposed a novel architecture, AASIST3, which enhances the original AASIST framework by incorporating Kolmogorov-Arnold networks, additional layers, and pre-emphasis. Furthermore, we introduced modifications using B-spline features as training features inspired by previous enhancements in synthetic speech detection models. In addition, we utilized additional data, scores fusion, and a self-supervised pre-trained model as an encoder to achieve the best results. Our findings indicated that these modifications significantly improve model performance, achieving a more than twofold improvement over AASIST. The model demonstrated minDCF results of 0.5357 under closed conditions and 0.1414 under open conditions, affirming the effectiveness of our configuration.

\newpage
\bibliographystyle{IEEEbib}
\bibliography{ASVspoof_BibEntries}

\begin{thebibliography}{10}

\bibitem{asvspoof13}
Nicholas Evans, Tomi Kinnunen, and Junichi Yamagishi,
\newblock ``Spoofing and countermeasures for automatic speaker verification,''
\newblock in {\em INTERSPEECH 2013, 14th Annual Conference of the International Speech Communication Association, August 25-29, 2013, Lyon, France}, ISCA, Ed., Lyon, 2013.

\bibitem{asvspoof15}
Zhizheng Wu et~al.,
\newblock ``Asvspoof 2015: the first automatic speaker verification spoofing and countermeasures challenge,''
\newblock in {\em INTERSPEECH 2015}, ISCA, Ed., Dresden, 2015.

\bibitem{asvspoof17}
Tomi Kinnunen, Md. Sahidullah, Héctor Delgado, et~al.,
\newblock ``{The ASVspoof 2017 Challenge: Assessing the Limits of Replay Spoofing Attack Detection},''
\newblock in {\em Proc. Interspeech 2017}, 2017, pp. 2--6.

\bibitem{asvspoof19}
Xin Wang, Junichi Yamagishi, Massimiliano Todisco, et~al.,
\newblock ``Asvspoof 2019: A large-scale public database of synthesized, converted and replayed speech,'' 2020.

\bibitem{asvspoof21}
Xuechen Liu et~al.,
\newblock ``Asvspoof 2021: Towards spoofed and deepfake speech detection in the wild,''
\newblock {\em IEEE/ACM Transactions on Audio, Speech, and Language Processing}, vol. 31, pp. 2507–2522, 2023.

\bibitem{singfake}
Yongyi Zang, You Zhang, Mojtaba Heydari, and Zhiyao Duan,
\newblock ``Singfake: Singing voice deepfake detection,'' 2024.

\bibitem{lavrentyeva19_interspeech}
Galina Lavrentyeva, Sergey Novoselov, Andzhukaev Tseren, Marina Volkova, Artem Gorlanov, and Alexandr Kozlov,
\newblock ``{STC Antispoofing Systems for the ASVspoof2019 Challenge},''
\newblock in {\em Proc. Interspeech 2019}, 2019, pp. 1033--1037.

\bibitem{choi22c_interspeech}
Sunmook Choi, Il-Youp Kwak, and Seungsang Oh,
\newblock ``{Overlapped Frequency-Distributed Network: Frequency-Aware Voice Spoofing Countermeasure},''
\newblock in {\em Proc. Interspeech 2022}, 2022, pp. 3558--3562.

\bibitem{alzantot19_interspeech}
Moustafa Alzantot, Ziqi Wang, and Mani~B. Srivastava,
\newblock ``{Deep Residual Neural Networks for Audio Spoofing Detection},''
\newblock in {\em Proc. Interspeech 2019}, 2019, pp. 1078--1082.

\bibitem{lai19b_interspeech}
Cheng-I Lai, Nanxin Chen, Jesús Villalba, and Najim Dehak,
\newblock ``{ASSERT: Anti-Spoofing with Squeeze-Excitation and Residual Networks},''
\newblock in {\em Proc. Interspeech 2019}, 2019, pp. 1013--1017.

\bibitem{castan22_odyssey}
Diego Castan et~al.,
\newblock ``{Speaker-Targeted Synthetic Speech Detection},''
\newblock in {\em Proc. The Speaker and Language Recognition Workshop (Odyssey 2022)}, 2022, pp. 62--69.

\bibitem{Kwak_artcle}
Il-Youp Kwak et~al.,
\newblock ``Voice spoofing detection through residual network, max feature map, and depthwise separable convolution,''
\newblock {\em IEEE Access}, vol. PP, pp. 1--1, 01 2023.

\bibitem{chen21_asvspoof}
Xinhui Chen, You Zhang, Ge~Zhu, and Zhiyao Duan,
\newblock ``{UR Channel-Robust Synthetic Speech Detection System for ASVspoof 2021},''
\newblock in {\em Proc. 2021 Edition of the Automatic Speaker Verification and Spoofing Countermeasures Challenge}, 2021, pp. 75--82.

\bibitem{Wu22}
Lei Wu and Ye~Jiang,
\newblock ``Attentional fusion tdnn for spoof speech detection,''
\newblock in {\em 2022 5th International Conference on Pattern Recognition and Artificial Intelligence (PRAI)}, 2022, pp. 651--657.

\bibitem{Khan23}
Awais Khan and Khalid Malik,
\newblock ``Spotnet: A spoofing-aware transformer network for effective synthetic speech detection,''
\newblock in {\em 2nd ACM International Workshop on Multimedia AI against Disinformation (MAD’23)}, 06 2023.

\bibitem{aasist}
Jee weon Jung et~al.,
\newblock ``Aasist: Audio anti-spoofing using integrated spectro-temporal graph attention networks,'' 2021.

\bibitem{aassist2}
Yuxiang Zhang, Jingze Lu, Zengqiang Shang, Wenchao Wang, and Pengyuan Zhang,
\newblock ``Improving short utterance anti-spoofing with aasist2,'' 2024.

\bibitem{w2v2_aasist}
Hemlata Tak et~al.,
\newblock ``Automatic speaker verification spoofing and deepfake detection using wav2vec 2.0 and data augmentation,'' 2022.

\bibitem{s2pecnet}
Penghui Wen et~al.,
\newblock ``{Robust Audio Anti-Spoofing with Fusion-Reconstruction Learning on Multi-Order Spectrograms},''
\newblock in {\em Proc. INTERSPEECH 2023}, 2023, pp. 271--275.

\bibitem{sam}
Pierre Foret, Ariel Kleiner, Hossein Mobahi, and Behnam Neyshabur,
\newblock ``Sharpness-aware minimization for efficiently improving generalization,'' 2021.

\bibitem{asam}
Jungmin Kwon, Jeongseop Kim, Hyunseo Park, and In~Kwon Choi,
\newblock ``Asam: Adaptive sharpness-aware minimization for scale-invariant learning of deep neural networks,'' 2021.

\bibitem{swl}
Zhiyong Wang, Ruibo Fu, Zhengqi Wen, Yuankun Xie, Yukun Liu, Xiaopeng Wang, Xuefei Liu, Yongwei Li, Jianhua Tao, Yi~Lu, Xin Qi, and Shuchen Shi,
\newblock ``Generalized fake audio detection via deep stable learning,'' 2024.

\bibitem{GCE}
Hye jin Shim, Md~Sahidullah, Jee weon Jung, Shinji Watanabe, and Tomi Kinnunen,
\newblock ``Beyond silence: Bias analysis through loss and asymmetric approach in audio anti-spoofing,'' 2024.

\bibitem{ding2022samospeakerattractormulticenter}
Siwen Ding, You Zhang, and Zhiyao Duan,
\newblock ``Samo: Speaker attractor multi-center one-class learning for voice anti-spoofing,'' 2022.

\bibitem{kan}
Ziming Liu, Yixuan Wang, Sachin Vaidya, Fabian Ruehle, James Halverson, Marin Soljačić, Thomas~Y. Hou, and Max Tegmark,
\newblock ``Kan: Kolmogorov-arnold networks,'' 2024.

\bibitem{kolmogorov1}
A.~N. Kolmogorov,
\newblock ``On the {Representation} of continuous functions of several variables as superpositions of continuous functions of a smaller number of variables.,''
\newblock {\em Dokl. Akad. Nauk}, vol. 108, no. 2, 1956.

\bibitem{kolmogorov2}
Jürgen Braun and Michael Griebel,
\newblock ``On a constructive proof of {Kolmogorov}’s superposition theorem,''
\newblock {\em Constructive approximation}, vol. 30, pp. 653--675, 2009,
\newblock Publisher: Springer.

\bibitem{kolmogorov3}
A.~N. Kolmogorov,
\newblock ``On the representation of continuous functions of many variables by superposition of continuous functions of one variable and addition,''
\newblock in {\em Doklady {Akademii} {Nauk}}. 1957, vol. 114, pp. 953--956, Russian Academy of Sciences.

\bibitem{sincconv}
Mirco Ravanelli and Yoshua Bengio,
\newblock ``Speaker recognition from raw waveform with sincnet,'' 2019.

\bibitem{w2v2}
Alexei Baevski, Henry Zhou, Abdelrahman Mohamed, and Michael Auli,
\newblock ``wav2vec 2.0: A framework for self-supervised learning of speech representations,'' 2020.

\bibitem{vaswani2017attention}
Ashish Vaswani, Noam Shazeer, Niki Parmar, Jakob Uszkoreit, Llion Jones, Aidan~N Gomez, {\L}ukasz Kaiser, and Illia Polosukhin,
\newblock ``Attention is all you need,''
\newblock {\em Advances in neural information processing systems}, vol. 30, 2017.

\bibitem{rawgat}
Hemlata Tak, Jee weon Jung, Jose Patino, Madhu Kamble, Massimiliano Todisco, and Nicholas Evans,
\newblock ``End-to-end spectro-temporal graph attention networks for speaker verification anti-spoofing and speech deepfake detection,'' 2021.

\bibitem{lin2017focal}
Tsung-Yi Lin, Priya Goyal, Ross Girshick, Kaiming He, and Piotr Doll{\'a}r,
\newblock ``Focal loss for dense object detection,''
\newblock in {\em Proceedings of the IEEE international conference on computer vision}, 2017, pp. 2980--2988.

\bibitem{yang2022algorithmic}
Tianbao Yang,
\newblock ``Algorithmic foundation of deep x-risk optimization,''
\newblock {\em arXiv preprint arXiv:2206.00439}, 2022.

\bibitem{yuan2023libauc}
Zhuoning Yuan, Dixian Zhu, Zi-Hao Qiu, Gang Li, Xuanhui Wang, and Tianbao Yang,
\newblock ``Libauc: A deep learning library for x-risk optimization,''
\newblock in {\em 29th SIGKDD Conference on Knowledge Discovery and Data Mining}, 2023.

\bibitem{res2netwithf0}
Cunhang Fan, Jun Xue, Jianhua Tao, Jiangyan Yi, Chenglong Wang, Chengshi Zheng, and Zhao Lv,
\newblock ``Spatial reconstructed local attention res2net with f0 subband for fake speech detection,'' 2023.

\bibitem{leaf}
Neil Zeghidour, Olivier Teboul, Félix de~Chaumont~Quitry, and Marco Tagliasacchi,
\newblock ``Leaf: A learnable frontend for audio classification,'' 2021.

\bibitem{chen2024towards}
William Chen et~al.,
\newblock ``Towards robust speech representation learning for thousands of languages,''
\newblock {\em arXiv preprint arXiv:2407.00837}, 2024.

\bibitem{attenaugmentconv}
Irwan Bello, Barret Zoph, Ashish Vaswani, Jonathon Shlens, and Quoc~V. Le,
\newblock ``Attention augmented convolutional networks,'' 2019.

\bibitem{rawboost}
Hemlata Tak, Madhu Kamble, Jose Patino, Massimiliano Todisco, and Nicholas Evans,
\newblock ``Rawboost: A raw data boosting and augmentation method applied to automatic speaker verification anti-spoofing,'' 2021.

\bibitem{arelu}
Dengsheng Chen, Jun Li, and Kai Xu,
\newblock ``Arelu: Attention-based rectified linear unit,'' 2020.

\bibitem{ukan}
Chenxin Li et~al.,
\newblock ``U-kan makes strong backbone for medical image segmentation and generation,'' 2024.

\bibitem{drokin2024kolmogorovarnoldconvolutionsdesignprinciples}
Ivan Drokin,
\newblock ``Kolmogorov-arnold convolutions: Design principles and empirical studies,'' 2024.

\bibitem{se}
Jie Hu, Li~Shen, Samuel Albanie, Gang Sun, and Enhua Wu,
\newblock ``Squeeze-and-excitation networks,'' 2017.

\bibitem{rawnet3}
Jee weon Jung, You~Jin Kim, Hee-Soo Heo, Bong-Jin Lee, Youngki Kwon, and Joon~Son Chung,
\newblock ``Pushing the limits of raw waveform speaker recognition,'' 2022.

\bibitem{wavenet}
Aaron van~den Oord, Sander Dieleman, Heiga Zen, Karen Simonyan, Oriol Vinyals, Alex Graves, Nal Kalchbrenner, Andrew Senior, and Koray Kavukcuoglu,
\newblock ``Wavenet: A generative model for raw audio,'' 2016.

\bibitem{ela}
Wei Xu and Yi~Wan,
\newblock ``Ela: Efficient local attention for deep convolutional neural networks,'' 2024.

\bibitem{arcface}
Jiankang Deng, Jia Guo, Jing Yang, Niannan Xue, Irene Kotsia, and Stefanos Zafeiriou,
\newblock ``Arcface: Additive angular margin loss for deep face recognition,'' 2018.

\bibitem{convnext}
Qiaowei Ma, Jinghui Zhong, Yitao Yang, Weiheng Liu, Ying Gao, and Wing~W.Y. Ng,
\newblock ``Convnext based neural network for audio anti-spoofing,'' 2022.

\end{thebibliography}

%

\end{document}